\documentclass[acmsmall, nonacm]{acmart}
\usepackage{algorithmic}
\usepackage{graphicx}
\usepackage{textcomp}
\usepackage{makecell}
\usepackage{subcaption}
\usepackage{listings}
\usepackage{fancyhdr}
\usepackage[dvipsnames]{xcolor}
\usepackage{multirow}
\usepackage{soul}
\definecolor{mygreen}{HTML}{2DA44E}
\usepackage{balance}
\usepackage{multicol}
\usepackage{xspace}
\usepackage{listings}
\usepackage{pifont}
\usepackage{xcolor}
\usepackage{threeparttable}
\usepackage{wrapfig}
\usepackage{tcolorbox}
\tcbuselibrary{listingsutf8}

\tcbset{
  colback=gray!10,
  colframe=black,
  listing only,
  listing options={basicstyle=\ttfamily\small,breaklines=true},
}

\usepackage{tcolorbox} 

\newcommand{\rqboxc}[1]{\begin{tcolorbox}[left=3pt,right=3pt,top=3pt,bottom=3pt,colback=gray!5,colframe=gray!40!black,before skip=5pt,after skip=5pt]#1\end{tcolorbox}}
\definecolor{brinkpink}{rgb}{0.98, 0.38, 0.5}

\newcommand{\peter}[1]{\textcolor{red}{{\it [Peter says: #1]}}}

\newcommand{\phead}[1]{\vspace{1mm} \noindent {\bf #1}}

\newcommand{\crs}{\emph{crash report}\xspace}

\newcommand{\Pagent}{\textit{Execution Analyzer Agent}\xspace}

\newcommand{\tooltable}{\textsc{Agentic-LLM}}

\newcommand{\devcrllm}{\textsc{Direct-LLM}}

\usepackage{pmboxdraw, amsmath}
\def\SFii{\textcolor{olive}{\textSFii\textSFx}}
\def\SFviii{\textcolor{olive}{\textSFviii\textSFx}}

\usepackage{booktabs}

\definecolor{dkgreen}{rgb}{0,0.6,0}
\definecolor{gray}{rgb}{0.5,0.5,0.5}
\definecolor{mauve}{rgb}{0.58,0,0.82}
\definecolor{darkgreen}{rgb}{0.01, 0.75, 0.24}
\AtBeginDocument{%
  }

\setcopyright{acmlicensed}
\copyrightyear{2018}
\acmYear{2018}
\acmDOI{XXXXXXX.XXXXXXX}
\acmConference{ACM International Conference}
\acmISBN{978-1-4503-XXXX-X/2018/06}

\begin{document}


\title{Crash Report Enhancement with Large Language Models: An Empirical Study}

\author{S M Farah Al Fahim}
\orcid{0009-0002-2187-2687}
\affiliation{%
  \institution{Software Performance, Analysis, and Reliability (SPEAR) Lab\\Concordia University}
  \city{Montreal}
  \country{Canada}
}
\email{smfarahal.fahim@mail.concordia.ca}

\author{Md Nakhla Rafi}
\orcid{0009-0005-4707-8985}
\affiliation{%
  \institution{Software Performance, Analysis, and Reliability (SPEAR) Lab\\Concordia University}
  \city{Montreal}
  \country{Canada}
}
\email{mdnakhla.rafi@mail.concordia.ca}

\author{Zeyang Ma}
\orcid{0000-0002-0390-1547}
\affiliation{%
  \institution{Software Performance, Analysis, and Reliability (SPEAR) Lab\\Concordia University}
  \city{Montreal}
  \country{Canada}
}
\email{m_zeyang@encs.concordia.ca}

\author{Dong Jae Kim}
\orcid{0000-0002-3181-0001}
\affiliation{%
  \institution{DePaul University}
  \city{Chicago}
  \country{USA}
}
\email{dkim121@depaul.edu}

\author{Tse-Hsun (Peter) Chen}
\orcid{0000-0003-4027-0905}
\affiliation{%
  \institution{Software Performance, Analysis, and Reliability (SPEAR) Lab\\Concordia University}
  \city{Montreal}
  \country{Canada}
}
\email{peterc@encs.concordia.ca}

\renewcommand{\shortauthors}{Fahim et al.}



\begin{abstract}
Crash reports are central to software maintenance, yet many lack the diagnostic detail developers need to debug efficiently. We examine whether large language models can enhance crash reports by adding fault locations, root-cause explanations, and repair suggestions. We study two enhancement strategies: \textsc{Direct-LLM}, a single-shot approach that uses stack-trace context, and \textsc{Agentic-LLM}, an iterative approach that explores the repository for additional evidence.
On a dataset of 492 real-world crash reports, LLM-enhanced reports improve Top-1 problem-localization accuracy from 10.6\% (original reports) to 40.2-43.1\%, and produce suggested fixes that closely resemble developer patches (CodeBLEU around 56-57\%). Both our manual evaluations and LLM-as-a-judge assessment show that \textsc{Agentic-LLM} delivers stronger root-cause explanations and more actionable repair guidance. A user study with 16 participants further confirms that enhanced reports make crashes easier to understand and resolve, with the largest improvement in repair guidance.
These results indicate that supplying LLMs with stack traces and repository code yields enhanced crash reports that are substantially more useful for debugging.
\end{abstract}



\maketitle

\section{Introduction}\label{sec:introduction}

Bug reports are essential artifacts in the software maintenance lifecycle. They help developers understand, localize, and resolve faults. Bettenburg et al.~\cite{bettenburg2008makes} found that high-quality bug reports typically include clear reproduction steps, information on affected components, expected versus observed behavior, and sometimes suggested fixes. However, in practice, developer-written bug reports are often incomplete or lack critical diagnostic context, making diagnosis difficult~\cite{bettenburg2008makes, rastkar2010summarizing}. Without clear and actionable information, developers must spend significant effort understanding the report, reproducing crashes, interpreting stack traces, or exploring the codebase to isolate issues. 


To assist developers, prior research has explored ways to enhance bug reports using execution artifacts and report structuring~\cite{nayrolles2015jcharming,dang2012rebucket,kim2013predicting,acharya2025can, derakhshanfar2020botsing}. 
Some approaches rely on stack traces to reproduce crashes or generate tests~\cite{9286108, nayrolles2015jcharming}, but these methods face challenges such as incomplete traces and difficulties in recovering the required inputs for large systems. Other works rewrite bug descriptions into structured templates, sometimes with the help of LLMs~\cite{acharya2025can}. Still, these approaches fall short because they cannot capture key source-code context, such as the execution paths that led to the crash. 
The recent advances of large language models (LLMs) provide an opportunity to go beyond simple text rewriting. By leveraging both the existing report content and repository-level source code context, LLMs can provide missing diagnostic details, such as likely faulty locations, root-cause explanations, and potential fixes. Such enhanced reports could significantly improve the usefulness of bug reports. Yet, it remains to be explored whether LLMs can reliably generate the diagnostic details required for effective debugging.

In this paper, we study the effectiveness of LLM, when combined with related source code information, on enhancing bug reports. In particular, we focus specifically on \textit{\textbf{crash reports}}, a structured subset of bug reports that include stack traces from runtime exceptions~\cite{chen2021pathidea}. Crash reports are particularly valuable because their stack traces capture concrete runtime execution paths, providing an explicit link to the source code and a natural starting point for diagnosis. Compared to more general bug reports, they are often more immediate to users and more directly connected to the underlying code. 

To investigate LLM-based crash report enhancement, we first construct an inter-procedural call graph by mapping each frame in the stack trace to its corresponding method in the repository source code. This call graph provides a program-level view of all potential execution paths that could lead to the crash. We then provide the LLM with both the original crash report and the source code context to generate an enhanced report that includes structured diagnostic fields for crash’s: \textit{Root Cause}, \textit{Steps To Reproduce}, \textit{Problem Location}, \textit{Repair Suggestion}, and \textit{Possible Fix}.

\begin{figure*}
    \centering
    \includegraphics[width=0.9\textwidth]{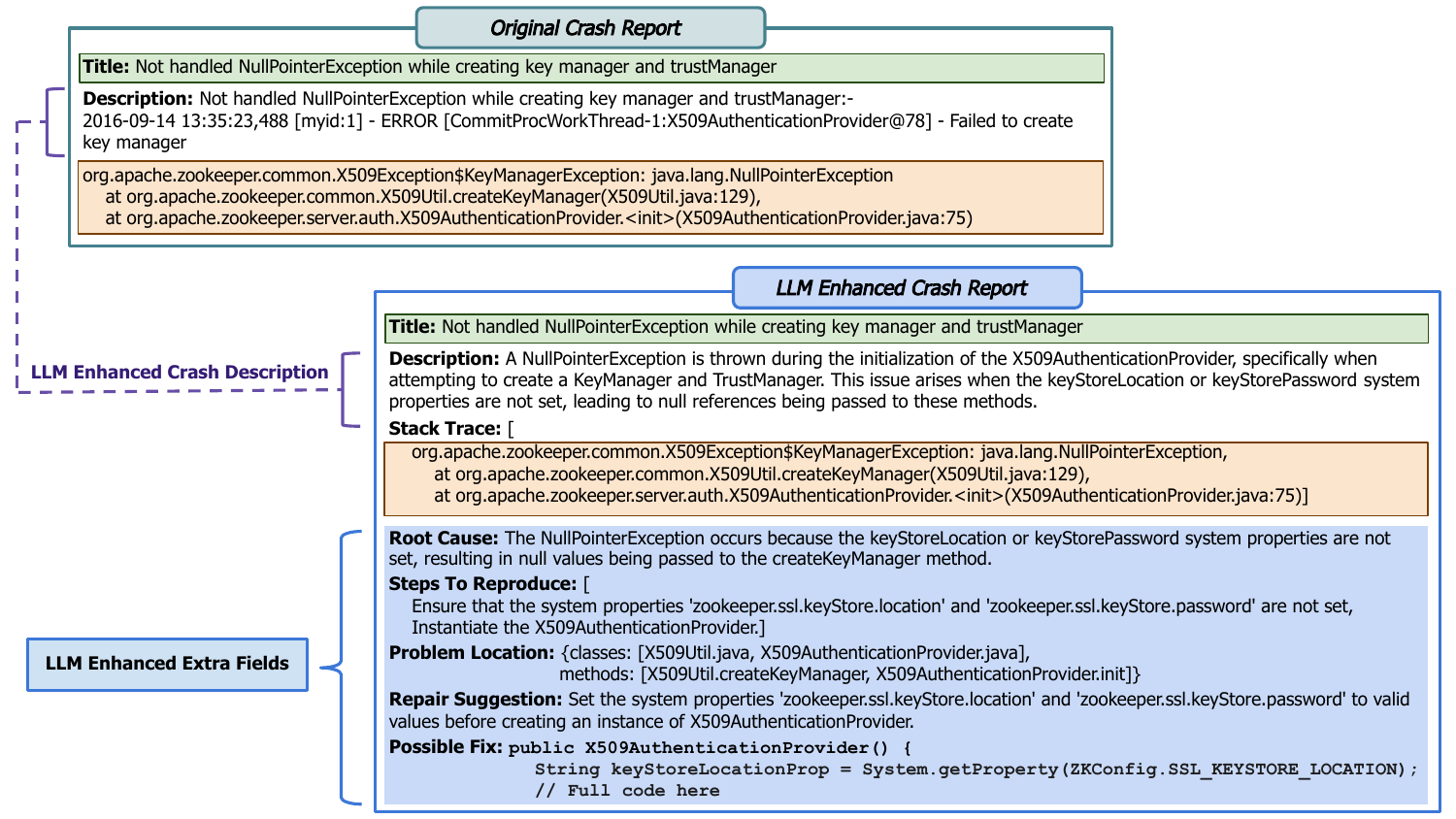}
    \caption{An end-to-end example showing the transformation of a developer-written crash report into a structured, LLM-enhanced report, based on the real-world issue ZOOKEEPER-2581.}
    \label{fig:original_vs_enhanced}
    \vspace{-1em}
\end{figure*}

We empirically study the enhancement through two instantiations: {\textsc{Direct-LLM}}, a single-shot strategy that augments a report using only the methods explicitly referenced in the stack trace, and {\textsc{Agentic-LLM}}, an iterative, agent-based strategy that traverses the call graph to collect additional evidence and synthesize structured diagnostic fields. Figure~\ref{fig:original_vs_enhanced} shows an example of the original crash report provided by developers and the report enhanced by \tooltable. Examining both instantiations on the same dataset and repository context allows us to isolate the effect of reasoning strategies. Namely, single-pass augmentation versus iterative exploration, and quantify their impact on the diagnostic information in the enhanced reports. 

We evaluate these two types of enhanced crash reports on a curated dataset of 492 real-world reports selected from eight widely used Java systems. Each crash report contains information on the corresponding source code and developer fixes. 
First, we perform a \textit{\textbf{quantitative evaluation}} to measure (i) the accuracy of problem localization (i.e., whether the reported faulty method is the same as the ground truth), and (ii) the similarity between the possible fixes and the developer patches using CodeBLEU. 
Second, we conduct a \textit{\textbf{qualitative evaluation}} by manually inspecting a stratified sample of the crash reports across three key dimensions: \textit{clarity of crash location}, \textit{completeness of crash's root cause}, and \textit{usefulness of repair suggestions}. 
Finally, we conduct a \textit{\textbf{user study}} with 16 participants to evaluate whether enhanced reports improve debugging compared to the original reports.

Our results show that LLM-enhanced crash reports consistently outperform developer-written reports across all dimensions. The enhanced reports' Top-1 problem localization accuracy improves fourfold compared to information retrieval-based baseline (from 10.6\% to 40.2–43.1\%), and the possible fixes achieve high CodeBLEU scores of 56\% with respect to developer patches. 
Our manual study and LLM-as-a-Judge evaluations find that the enhanced reports provide clearer crash locations, more complete root-cause explanations, and higher-quality repair guidance. 
Moreover, participants in our user study found LLM-enhanced reports substantially more helpful for understanding and resolving crashes, giving them an average rating of 4.59/5, compared to 3.0/5 for the original reports.
In short, our findings provide strong empirical evidence on the effectiveness of leveraging LLM and source code context for enhancing crash reports to assist developers in problem diagnosis. 

Our contributions are summarized as follows: 

\begin{itemize}
    \item To the best of our knowledge, this is the first empirical study that combines runtime execution signals (stack traces) with repository-level source code to enhance crash reports.

    \item We design two instantiations to study the impact of reasoning strategy: \textsc{Direct-LLM}, a single-shot approach that enhances reports using only stack trace referenced code, and \textsc{Agentic-LLM}, an agentic approach that autonomously traverses the call graph and enhances the report through iterative reasoning. Both approaches significantly improve crash report quality, but \textsc{Agentic-LLM} yields more comprehensive diagnostics at a higher computational cost (\$0.01 per crash report vs \$0.005).

    \item Using a curated dataset of 492 real-world crash reports from eight widely used Java projects, we show that LLM-enhanced reports substantially improve problem localization accuracy (up to $4\times$), generate high-quality repair suggestions, and outperform developer-written reports on diagnostic quality.

    \item  A study with 16 users shows that participants find LLM-enhanced reports significantly more helpful for understanding and resolving crashes.

    \item We release our curated dataset, enhancement pipeline, and evaluation artifacts to support reproducibility and future research.

\end{itemize}

Our findings highlight the potential of leveraging execution context with LLM to generate actionable and higher-quality crash reports. This work opens a promising direction for developing automated debugging workflows that reduce developer effort and accelerate fault resolution.  

\phead{Paper Organization.} Section~\ref{sec:background} discusses background and related work. Section~\ref{sec:empirical} presents data selection \& preliminary analysis. Section~\ref{sec:approach} presents study design. Section~\ref{sec:discussion} reports experimental results. Section~\ref{sec:threats} discusses threats to validity. Finally, Section~\ref{sec:conclusion} concludes the paper.

\section{Motivation and Related Work}\label{sec:background}

 \begin{figure*}[h!]
   \centering
   \includegraphics[width=0.9\textwidth]{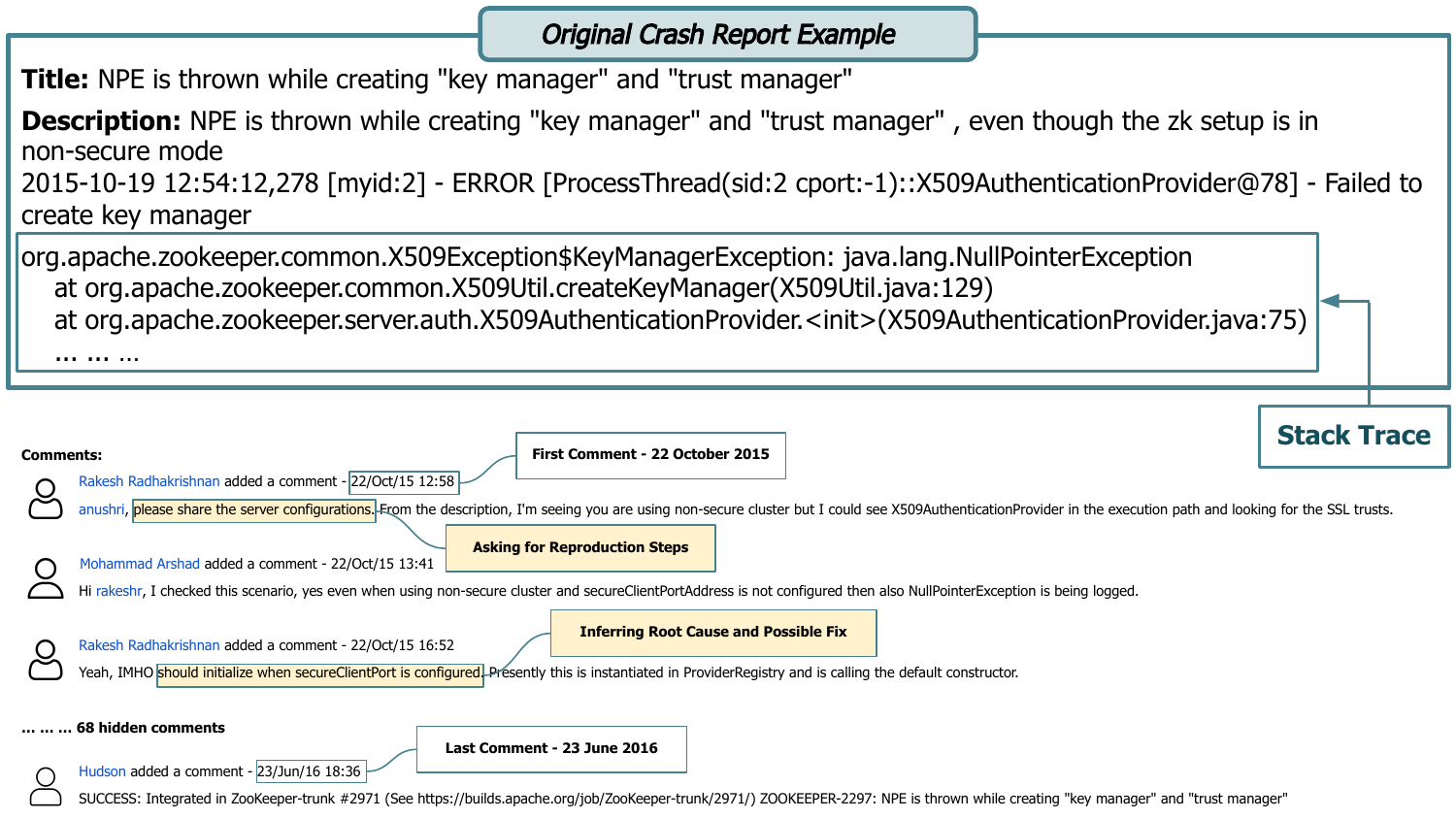}
   \caption{Raw description field and developer comments for bug ZOOKEEPER-2297. The report omits reproduction steps, configuration details, and suggested fixes, requiring developers to infer missing information through discussion.}
   \label{fig:bug-report-example}
   \vspace{-1em}
 \end{figure*}

\subsection{Quality Issues in Crash Report}
Developers frequently rely on crash reports to identify the root cause of bugs. In common issue trackers, a crash report typically consists of a \emph{title}, a \emph{description}, and an embedded \emph{stack trace} that captures the runtime state at the moment of failure~\cite{bettenburg2008makes, summarize2010rastakar, ahmed2014impact}. Prior studies highlight that high-quality reports provide not only stack traces but also clear reproduction steps and supporting contextual information, which substantially accelerate bug triage and resolution~\cite{bettenburg2008makes, summarize2010rastakar, hooimeijer2007modeling}. Such quality is especially critical for enabling downstream automated debugging techniques, including automatic fault localization~\cite{gong2014locating} and automated program repair~\cite{monperrus2018automatic}.

Despite their importance, the quality and completeness of 
\crs vary widely in practice~\cite{bettenburg2008makes, rastkar2010summarizing}. To illustrate, Figure~\ref{fig:bug-report-example} presents a report from the Apache ZooKeeper project (ZOOKEEPER-2297). This report includes only a short title and a minimal description field. The description contains a single sentence nearly identical to the title, followed by a stack trace, without reproduction steps, affected software components, or a comparison of expected versus observed behavior. 
For instance, one developer asked, \textit{"Please share the server configurations"}, indicating that reproduction steps and environment details were missing. Another remarked \textit{"should initialize when secureClientPort is configured"}, reflecting that both the root cause and a possible fix had to be inferred during discussion rather than being provided in the report.
As a result, developers engaged in a prolonged back-and-forth discussion to request missing information, and the issue remained unresolved for months. The issue was reported on \textit{October 19, 2015}, and was not resolved until \textit{June 23, 2016}, a process that took approximately \textit{eight months}. This highlights the challenges of incomplete reports in fault diagnosis: while crash reports are essential, they often lack sufficient context, forcing developers to perform root cause analysis manually, a time-consuming process. 

These challenges raise a central question: \textit{{how can we make crash reports more complete and diagnostically useful to improve bug triage for developers?}} Traditional approaches to enhancing crash reports, such as rule-based heuristics~\cite{qian2023survey}, or manual instrumentation~\cite{wang2016improving}, often require substantial developer effort~\cite{zou2018practitioners}, and may fail to capture semantic relationships between program state and failure context. This limitation motivates exploring alternative approaches that can reason over both code and natural language. Large Language Models (LLMs), pre-trained on extensive corpora of natural language and source code, offer a promising alternative. LLMs have demonstrated impressive capabilities across a wide range of tasks in natural language processing~\cite{brown2020language, achiam2023gpt} and software engineering domains~\cite{chen2021evaluating, vaithilingam2022expectation, lu2023llama}. Their performance, coupled with the growing availability of powerful open-source models and recent breakthroughs in agentic capability~\cite{liu2023agentbench, hong2023metagpt, xu2023exploring, qian2023chatdev, lin2024soen}, raises the possibility of reducing developer effort, particularly in enhancing crash reports. Our research aims to show that LLMs can be utilized to generate more complete diagnostic crash reports.




\subsection{Related Work}\label{sec:related}

Bug reports, and especially crash reports that include stack traces, play a vital role in debugging and fault localization. We organize the related work as follows: first, prior work that most directly addresses generating or improving developer-written bug/crash reports; second, literature on fault localization and diagnostics that motivates using stack traces and source-code context as inputs.


\phead{Bug report generation and enhancement.}
Bug reports, especially those with stack traces, are essential for debugging. Earlier work has shown that execution artifacts, such as stack traces, can help reproduce crashes and prioritize reports. 
For example, JCHARMING~\cite{nayrolles2015jcharming} follows stack traces and applies directed model checking to reproduce Java crashes, but still struggles with scalability to large systems, incomplete traces, or environment-dependent failures.
ReBucket~\cite{dang2012rebucket} groups duplicate reports using call-stack similarity to help triage, and Kim et al.~\cite{kim2013predicting} show that stack traces can be used to predict crash-prone methods. 
More recently, large language models (LLMs) have been applied to bug reports. Acharya and Ginde~\cite{acharya2025can} use LLMs to convert unstructured bug descriptions into structured templates (e.g., Steps, Expected/Actual behavior). However, their approach only involves rewriting the existing text in the report, without including additional context, such as the source code and the execution paths that led to the issue. 
To the best of our knowledge, existing work has not examined whether LLMs can enhance crash reports by combining stack traces and source-code context. We address this gap with an empirical study on such artifact-grounded crash-report enhancement.

\phead{Fault localization and diagnostics.}
Prior work shows that stack traces and other program artifacts help find and reproduce faults. For example, BLUiR~\cite{saha2013improving} uses structured information from stack traces and program analysis to link reports to relevant code components. BRTracer~\cite{wong2014boosting} proposes methods to enhance fault localization by analyzing bug reports and stack traces, including segmentation and stack trace analysis techniques.

More recently, several LLM-based studies have explored how models can assist localization and provide diagnostics. AutoFL~\cite{kang2024quantitative} equips models with repository-navigation tools so they can fetch code snippets and produce ranked suspects together with natural-language explanations. LLMAO~\cite{yang2024large} trains a small task-specific component on top of a pretrained language model to assign suspiciousness scores to code lines, enabling fault localization without executing tests. Wu et al.\cite{wu2023large} empirically evaluated ChatGPT models against traditional techniques on the Defects4J benchmark~\cite{defects4j}. AgentFL~\cite{qin2024agentfl} decomposes localization into comprehension, navigation, and confirmation steps using multiple specialized agents. 
LLM4FL~\cite{rafi2024multi} introduces a multi-agent fault localization framework that addresses LLMs' token limitations by splitting coverage data and enhancing repository-level analysis through graph-based code retrieval. It then refines the ranking of suspicious methods via iterative self-reflection.

These prior works directly informed our design choices, for example, which report fields to extract, how much code context to provide, and how to structure agentic workflows. At the same time, their goals and outputs differ from ours: they typically produce ranked suspects, test scripts, or explanatory text, whereas we empirically evaluate LLM enhancement for crash reports with stack traces and source code context.

\section{Data Selection \& Preliminary Analysis}
\label{sec:empirical}

Our research goal is to show that LLM-enhanced crash reports can provide complete diagnostic information for better developer bug triage. In this section, we present a preliminary study on existing quality issues in developer-written crash reports.


\subsection{Dataset Selection \& Refinement}
\label{sec:dataset}

\begin{wraptable}{r}{0.5\textwidth} 
\caption{An overview of our studied systems. LOC reports the LOC of the version captured in the Pathidea snapshot~\cite{chen2021pathidea}.}
\centering
\scalebox{.7}{
\setlength{\tabcolsep}{0.9em}
\begin{tabular}{lcc}
    \toprule
    \textbf{System} & \textbf{LOC} & \textbf{\# Crash Reports} \\
    \midrule
    ActiveMQ         & 338k   & 35  \\
    Hadoop Common    & 190k  & 38  \\
    HDFS             & 285k  & 73  \\
    MapReduce        & 198k  & 65  \\
    YARN             & 548k  & 117 \\
    Hive             & 1.2M  & 120 \\
    Storm            & 275k   & 36  \\
    ZooKeeper        & 79k   & 8  \\
    \midrule
    \textbf{Total}   & \textbf{3.1M} & \textbf{492} \\
    \bottomrule
\end{tabular}
}
\label{tab:pathidea_data}
\end{wraptable}
We build on the Pathidea dataset~\cite{chen2021pathidea}, which includes real-world Jira bug reports with logs collected from eight widely-used, Java-based open-source systems (see Table~\ref{tab:pathidea_data}). We focus exclusively on \emph{crash reports}, i.e., bug reports that include a stack trace capturing runtime failures. Such reports are valuable artifacts for debugging because stack traces provide concrete execution paths and an explicit link to the source code, offering a natural starting point for diagnosis. In addition to stack traces, crash reports may also include developer-written textual descriptions and system logs, further enriching the information available for analysis. Table~\ref{tab:pathidea_data} summarizes the eight open-source systems in our analysis, listing the number of extracted crash reports and the LOC of each system at the Pathidea snapshot~\cite{chen2021pathidea}.

We applied a series of filtering and refinement steps to select the bug reports for our study. First, to ensure each bug report corresponds to a confirmed and actionable issue (i.e., one that led to a verified fix), we retained only those labeled as \texttt{Resolved} or \texttt{Fixed} and verified that they have associated commits in the version control history. This ensures traceability between the report and the actual fix, which is important for evaluating tasks such as fault localization~\cite{kang2024quantitative,rafi2024multi,li2021fault} and bug explanation. Next, we restricted the dataset to high-priority issues (\texttt{Major}, \texttt{Critical}, or \texttt{Blocker}), as these are more likely to correspond to impactful failures. We limited reports to those created in 2010 or later to align with modern development practices and tooling.

Finally, from this filtered pool of bug reports, we then extracted the subset that qualify as \emph{crash reports}. We used regular-expression matching on the \texttt{description} field to detect stack trace patterns, yielding 492 reports that form our final dataset. From each \emph{crash report}, we retained fields most relevant for debugging, such as \texttt{title} and \texttt{description}, while discarding metadata including assignee, timestamps, affected versions, and custom fields~\cite{wang2016amalgam+}. 

\subsection{Preliminary Analysis on the Quality of the Crash Reports}
\label{sec:preliminary_analysis}


Bettenburg et al.~\cite{bettenburg2008makes} suggest that a high-quality crash report often contains: (1) \emph{steps to reproduce}, (2) an explicit \emph{root-cause hypothesis}, (3) \emph{affected component(s) or file(s)} (e.g., classes/modules), and (4) a \emph{suggested fix} (textual guidance and/or a code change). 
To quantify the frequency with which original reports include these attributes, we conducted a manual examination of all 492 crash reports. Since the original crash reports did not consistently provide dedicated fields for these attributes, we examined only the \texttt{title} and \texttt{description} fields when searching for them. Two authors independently annotated each report for the presence or absence of the four attributes, and no disagreements were observed.

Table~\ref{tab:presence_audit} summarizes the results of manual analysis. The numbers reveal a substantial gap between diagnostic attributes that best support debugging and information that developers typically provide. 
Mentions of affected components were present in only 187 reports (38.01\%), and explicit root-cause hypotheses in 196 reports (39.84\%). In contrast, reproduction steps appeared in just 58 reports (11.79\%) and suggested fixes in 103 reports (20.93\%). These results reinforce prior observations about quality issues in crash reports~\cite{bettenburg2008makes, rastkar2010summarizing}. \textit{\textbf{While crash reports often include stack traces that capture fatal runtime failures, many provide little beyond the trace itself, omitting critical information such as reproduction steps and root-cause analysis}}.

\begin{wraptable}{r}{0.55\textwidth} 
\caption{The presence of each important attribute~\cite{bettenburg2008makes} in the studied 492 crash reports.}
\vspace{-1em} 
\centering
\scalebox{.8}{
\setlength{\tabcolsep}{1.1mm}
\begin{tabular}{lccc}
\toprule
\textbf{Attribute} & \textbf{Present} & \textbf{Absent} & \textbf{\% Present} \\
\midrule
Steps to reproduce & {58} & {434} & {11.79\%} \\
Root-cause hypothesis & {196} & {296} & {39.84\%} \\
Affected component(s)/file(s) & {187} & {305} & {38.01\%} \\
Suggested fix & {103} & {389} & {20.93\%} \\
\bottomrule
\end{tabular}
}
\label{tab:presence_audit}
\end{wraptable}

\section{Study Design}\label{sec:approach}



Original developer-provided crash reports frequently omit diagnostic information, which hinders bug triage and diagnosis. To address this limitation, we conduct an empirical study on the usefulness of LLM-based crash report enhancement through repository-level analysis. Our goal is to examine whether LLM can help improve report completeness and support more effective triage.

To evaluate the effectiveness of LLMs for crash-report enhancement, we empirically compare two instantiations: \textsc{Direct-LLM} and \textsc{Agentic-LLM}. Both instantiations rely on the same preprocessed repository artifacts, stack traces, a checked-out repository snapshot, and an inter-procedural call graph (representing the potential execution paths when the exception occurred). 
Although both instantiations can have access to the same preprocessed data, they differ in how they retrieve and expose those artifacts to the model at runtime as inputs. \textsc{Direct-LLM} is a single-shot invocation that uses only the original crash report and the method bodies explicitly referenced in the stack trace, without exploring additional callers-callees; inputs that exceed the context window are truncated as described in Section~\ref{sec:direct_llm}. In contrast, \textsc{Agentic-LLM} employs an iterative, multi-agent workflow that traverses repository artifacts, targeting complex bug triage scenarios that require multi-step reasoning (e.g., root cause analysis). Studying both workflows enables us to assess the trade-offs between efficiency and LLM-trajectory depth, highlighting when lightweight, single-shot reasoning is sufficient and when agentic orchestration is necessary for comprehensive crash-report enhancement.

 \begin{figure*}
   \centering
   \includegraphics[width=0.9\textwidth]{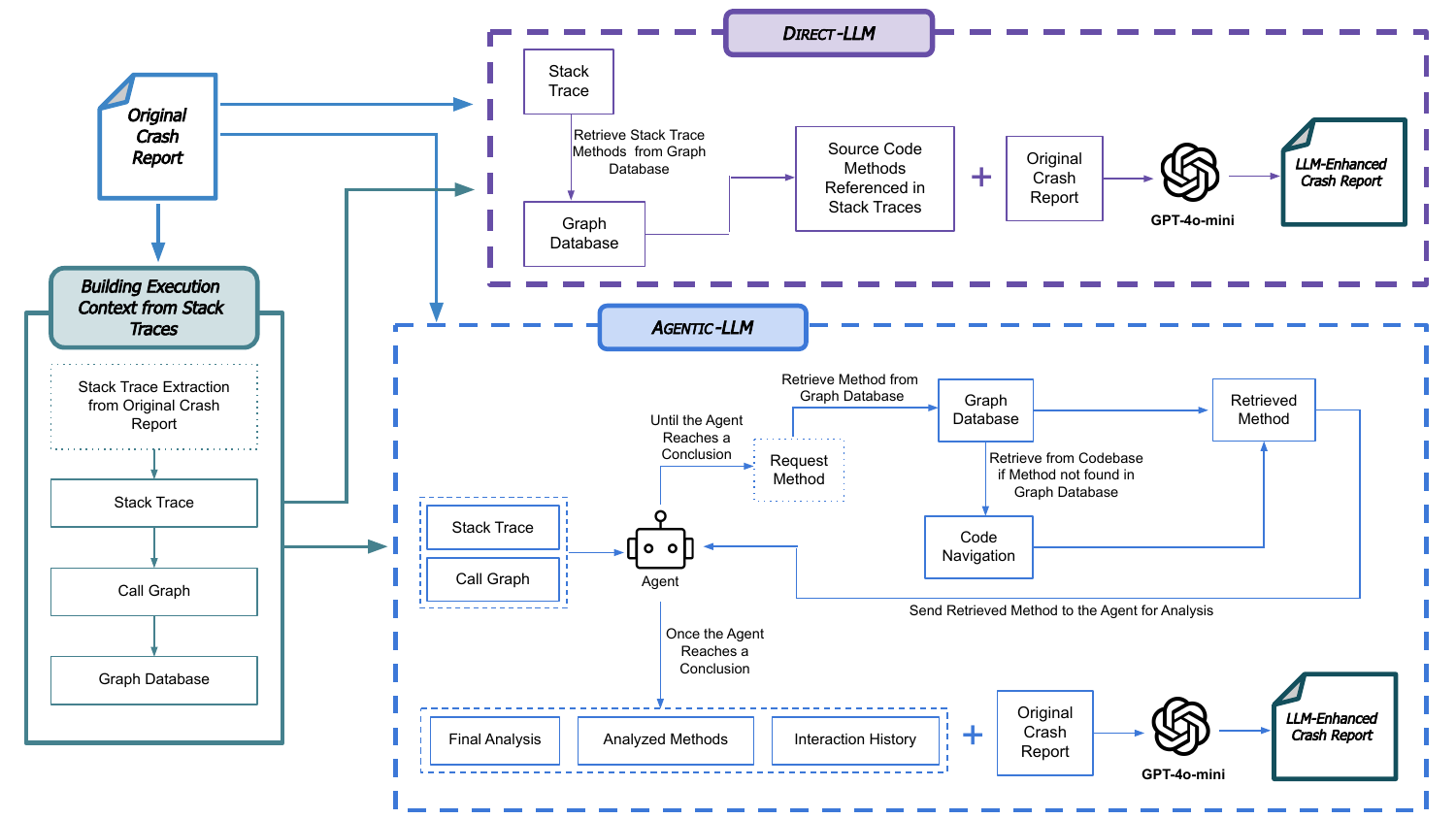}
   \caption{An overview of our study. \textsc{Direct-LLM} performs single-shot enhancement, while \textsc{Agentic-LLM} performs iterative enhancement. Both analyze crash reports by reasoning over repository-level artifacts obtained through preprocessing and source code indexing, but they differ in their overall workflow strategy.} 
   \label{fig:overview}
   \vspace{-1em}
 \end{figure*}

Figure~\ref{fig:overview} illustrates the overall pipeline of our study. Starting from an original \textit{crash report}, we extract its stack trace, check out the corresponding repository snapshot, parse the source code to build an inter-procedural call graph, and construct a method-level graph database linking method signatures to their full bodies. From this shared execution context, the two LLM-based instantiations diverge. \textsc{Direct-LLM} fetches only the methods explicitly referenced in the stack trace, combines those method bodies with the original report, and invokes a single-shot, meta-reasoning prompt~\cite{gao2024meta} to produce its enhanced crash report. \textsc{Agentic-LLM} employs \textit{Execution Path Analyzer} agent that starts from the stack trace, iteratively requests methods from the graph database (falling back to the full codebase when needed), reasons over retrieved methods while maintaining an interaction history, and repeats until a conclusion.


\subsection{Building Execution Context from Stack Traces}
\label{sec:shared_preproc}

Both \textsc{Direct-LLM} and \textsc{Agentic-LLM} analyze the crash report by reasoning over the repository-level artifacts obtained through preprocessing and source code indexing. For each crash report, we preprocess the repository-level context by (i) extracting the stack traces, (ii) checking out the corresponding repository version based on the crash report creation time, and (iii) constructing a static inter-procedural call graph to re-construct the execution paths and enable method-level retrieval. The two workflows differ only in workflow strategy.

\phead{Repository Checkout for Crash Analysis.} 
For each crash, we identify the repository version corresponding to the reported crash time. Using \texttt{creation\_time} metadata from the Jira crash report, we determine the commit just before the crash was reported, corresponding to the crash. We then stash the untracked changes, pull the latest updates from the main branch, and check out the identified commit. This helps ensure that static analysis is performed on the correct historical snapshot of the repository.


\phead{Stack Trace Extraction from Unstructured Crash Reports.} 
We extract stack traces from the crash-report description using a regex-based extractor. We locate an exception header (a line containing "\texttt{Exception}", "\texttt{Error}", or "\texttt{Caused by:}") and capture the subsequent lines that begin with "\texttt{at}" and reference a Java source file (filename ending with "\texttt{.java}"). The extracted trace is represented as: 
\(
S = \{m_1, m_2, \dots, m_n\}.
\)

\phead{Augmenting Stack Traces with Static Call Graph.}
While stack traces reveal the sequence of method calls at the point of crash, they omit many feasible execution paths that could contribute to the failure.
To fill in the missing paths, we augment the stack trace with a static inter-procedural call graph derived from the source code.
We parse the Java source code using \texttt{javalang}~\cite{javalang} to build abstract syntax trees (ASTs) for each file, and we extract the method declarations and identify invocation sites. From this, we construct a directed call graph:
$G = (V, E)$, 
where each node \( v \in V \) represents a fully qualified method, and each edge \( (v_i, v_j) \in E \) indicates a direct method invocation from \( v_i \) to \( v_j \). Starting from the methods observed in the stack trace, we recursively traverse $G$ to capture all methods that could have been executed along the potential call paths leading to the crash. The graph $G$ is stored in a graph database that supports fast neighbor queries and method-source lookup. For example, given a method signature $v_j$, both \textsc{Direct-LLM} and \textsc{Agentic-LLM} can retrieve source code context from the graph database without re-parsing source files.

\subsection{LLM-based Crash Report Enhancement}

\subsubsection{\textsc{\textbf{Direct-LLM}}}
\label{sec:direct_llm}

\begin{wrapfigure}{r}{0.45\textwidth} 
  \centering
  \includegraphics[width=0.45\textwidth]{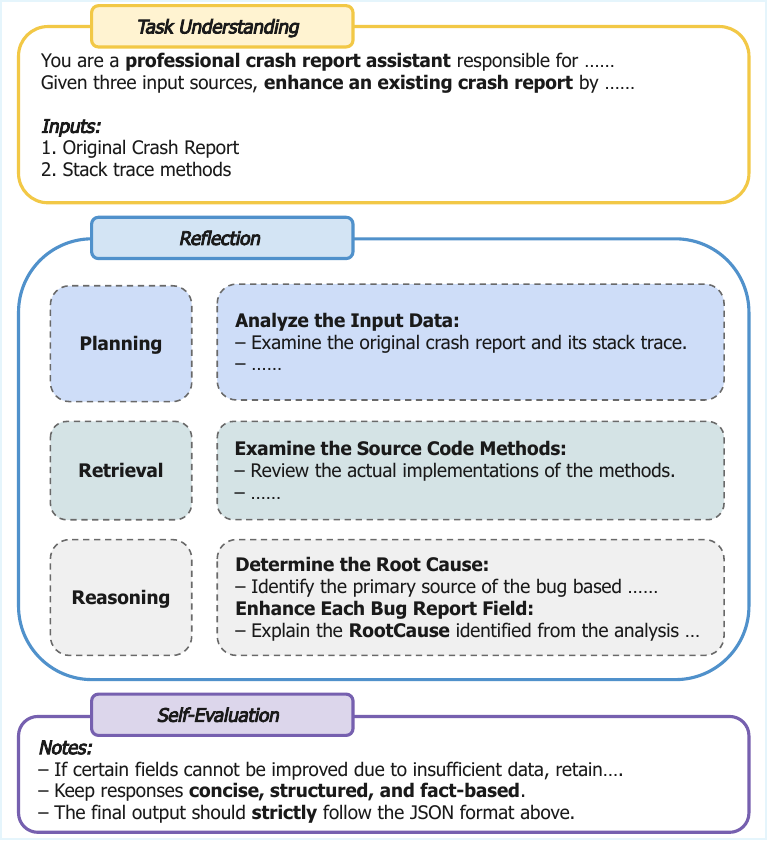}
     \caption{Meta-reasoning prompt for crash report enhancement. This prompt structures the LLM’s reasoning into five stages (task understanding, planning, retrieval, reasoning, and self-evaluation).}
  \label{fig:mrp_report_generator}
\end{wrapfigure}

The \textsc{Direct-LLM} performs a single-shot enhancement of each crash report. For this purpose, LLM is provided with three contexts: (i) the original crash report (title and description), (ii) the preprocessed stack trace \(S\) (Section~\ref{sec:shared_preproc}), and (iii) the source code of the methods referenced by the stack trace \(S\). Each parsed frame in \(S\) is indexed by its corresponding method node in the graph database, from which we retrieve the complete method body. Notably, \textsc{Direct-LLM} does not traverse to callers or callees beyond those explicitly present in \(S\). In cases where the input exceeds the model’s context window, we maintain the original stack-frame order and apply truncation from the earliest frames (i.e., those farthest from the thrown exception), thereby prioritizing frames most proximal to the exception.

To ensure consistent, structured reasoning, we adopt Meta-Reasoning Prompting (MRP) style to enhance the crash report~\cite{wang2023metacognitive}. Rather than prompting free-form instructions, which may produce inconsistent output, the prompt is designed to guide the LLM through a sequence of reasoning stages: task comprehension, planning, retrieval, reasoning, and self-evaluation. Prior work shows that such meta-reasoning frameworks improve accuracy, robustness, and consistency compared to free-form instructions~\cite{gao2024meta, wang2023metacognitive, zhang2023meta, suzgun2024meta}. By meta-prompting, we ensure crash report enhancements are well structured and align with the context provided.


Figure~\ref{fig:mrp_report_generator} shows the MRP-guided prompt for crash report enhancement. It is organized into five key fields. In the \textit{\ding{172}  task understanding} phase, it is assigned the role of a \textit{“professional crash report assistant”}, with clearly defined goals and input expectations. The \textit{\ding{173} planning} stage instructs it to follow a structured approach: parse inputs, analyze relevant methods, and synthesize outputs. In the \textit{\ding{174} retrieval} stage, it integrates context from the stack trace \( S \) and stack trace methods. During the \textit{\ding{175} reasoning} phase, it connects symptoms with observed failures and source-level logic. Finally, the \textit{\ding{176} self-evaluation} phase ensures that the generated content is consistent and grounded in the available evidence. 

\subsubsection{\textsc{\textbf{Agentic LLM}}}
\label{sec:agentic_llm}

The \textsc{Agentic-LLM} augments the execution-context from Section~\ref{sec:shared_preproc} (e.g., stack trace \(S\), checked-out repository snapshot, method-level index, and call graph \(G\)) by coordinating two tools: (1) an \textbf{Execution Path Analyzer} that incrementally retrieves and reasons about methods along the call graph using a GraphRAG-based approach (we denote this component \Pagent), and (2) a final report-generation step that generates the final structured report from the evidence gathered by the analyzer. This design trades additional retrieval and inference cost for deeper reasoning of the repository code.

\begin{wrapfigure}{r}{0.45\textwidth} 
\vspace{-0.3cm}
  \centering
  \includegraphics[width=0.45\textwidth]{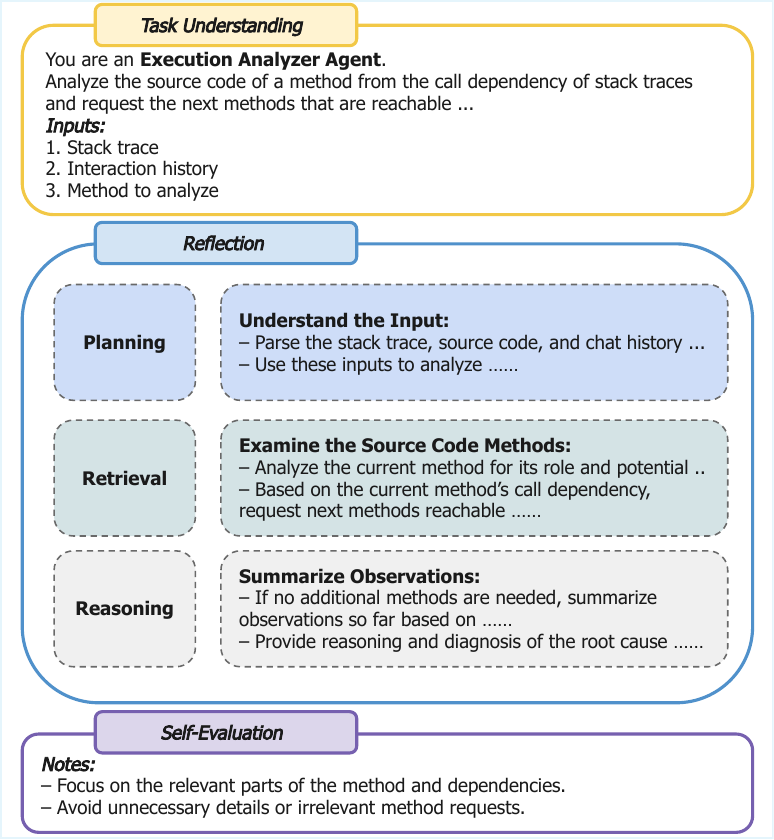}
  \caption{Meta-reasoning prompt for iterative context retrieval in \textbf\Pagent.}
  \label{fig:mrp_analyzer_agent}
\end{wrapfigure}

Both \textsc{Direct-LLM} and \textsc{Agentic-LLM} rely on the meta-reasoning prompting technique (Figure~\ref{fig:mrp_report_generator}) to guide the structured generation process, with their respective contexts supplied through the workflow. The key difference is that \textsc{Direct-LLM} proceeds to this step directly, operating on the original crash report and stack trace methods only. In contrast, \textsc{Agentic-LLM} inserts an additional iterative phase (Figure~\ref{fig:mrp_analyzer_agent}), where \Pagent iteratively retrieves relevant methods and code context.


\paragraph{\textbf{\Pagent}}
\label{sec:pagent}
Repository-level analysis is challenging for LLMs due to limited context sizes and the need to reason across large codebases~\cite{liu2024repoqa, rafi2024multi}. This challenge is further amplified when diagnosing software crashes: stack traces expose only partial execution paths, requiring the LLM to recover missing context from the repository. To retrieve the missing paths beyond stack traces, the \Pagent adopts GraphRAG, a graph-based retrieval technique~\cite{edge2024local} that incrementally fetches the caller-callee methods using the pre-constructed call graph (Section~\ref{sec:shared_preproc}). In the following, we explain the design of \Pagent and how it reconstructs the execution context for crash report enhancement.  



\phead{GraphRAG-Guided Execution Path Traversal for Repository-Level Diagnosis.} 
 \Pagent uses the stack trace \(S\) and an inter-procedural call graph \( \mathcal{G} = (\mathcal{V}, \mathcal{E}) \) to iteratively traverse the graph, re-constructing a more complete view of the program’s execution and uncovering methods that may cause the observed crash. This analysis starts from the top-most frame in the stack trace, the method where the exception was thrown, and incrementally explores the frame to traverse methods along the execution path.

At each step, the agent selects a method \( m \in \mathcal{V} \) to analyze and invokes the \texttt{provide\_method(m)} tool, responsible for providing the source code method to the agent, to retrieve its source code. Since loading the entire codebase into an LLM is infeasible, we adopt a \textit{GraphRAG-based retrieval} strategy to resolve only the relevant parts of the repository incrementally. In particular, this retrieval process prioritizes methods directly connected to \( m \) in the call graph (i.e., callers and callees) and tracks previously visited methods to avoid recursive loops or redundant analysis.

\phead{Iterative Fault Reasoning over Execution Paths.} Agent uses the MRP-guided prompt shown in Figure~\ref{fig:mrp_analyzer_agent}. At each step, the agent retrieves a method from the call graph, analyzes its body in combination with the stack trace \(S\) and the accumulated interaction history \(H\). The prompt guides the agent to decide whether there is sufficient context for root cause explanation. If analysis is inconclusive, it updates the candidate set \(C_t\), requests the next caller or callee from the call graph, and logs the action to \(H\). This traversal loop continues until the agent identifies a conclusive explanation for the failure, exhausts the reachable methods, or determines that further reasoning is unlikely to yield additional insight. To ensure scalability, the call-graph database is built once, retrieves only neighborhood methods (callers and callees) rather than entire files, and caches prior reasoning for reuse. Together, these strategies reduce both token usage and LLM calls and also enable more efficient repository analysis.

\phead{Final Enhanced Crash Report Generation.}
\textsc{Agentic-LLM} enhances the crash report using the same MRP-structured prompt as shown in Figure~\ref{fig:mrp_report_generator}, drawing context from the \textit{agentic context search}: the original crash report, the set of analyzed methods \(\mathcal{M}^*\), and the interaction history \(H\). The prompt then generates the final enhanced report in JSON structured format. 
Overall, \textsc{Agentic-LLM} coordinates a modular pipeline: the preprocessing step builds the execution context (stack trace, repository snapshot, method index, and call graph), the \Pagent iteratively investigates the root cause by traversing and reasoning over relevant methods, and the final prompt produces a structured and enhanced crash report. This design trades additional retrieval and inference cost for deeper, stepwise reasoning of the repository code.

\subsection{An End-to-End Example}
\label{sec:end_to_end_example}

To illustrate how LLMs transform an under-specified crash report into a structured and actionable diagnostic artifact, we present a real-world failure from the Apache ZooKeeper project (ZOOKEEPER-2581). As shown in Figure~\ref{fig:original_vs_enhanced}, the original report includes a title and a description field that merely repeats the title, followed by a log message and the raw stack trace. It does not explain the failure, provide no reproduction steps, and offers no insight into the root cause, leaving developers with minimal context to understand or debug the issue.

The \emph{LLM-enhanced} report shown here is produced by the \textsc{Agentic-LLM} instantiation and is presented as a representative end-to-end example because it exercises the full iterative retrieval and reasoning pipeline (GraphRAG and Execution Path Analyzer). 
Starting from the same raw inputs, our preprocessing extracts the stack trace and builds the repository snapshot and call graph. The \Pagent (in the \textsc{Agentic-LLM} instantiation) then selectively traverses the execution neighborhood using GraphRAG to inspect methods implicated in the failure. Through this process, the analyzer uncovers that the \texttt{X509AuthenticationProvider} constructor attempts to initialize a KeyManager and TrustManager using utility methods in \texttt{X509Util}, but fails due to missing system properties. The corresponding methods \texttt{createKeyManager} and \texttt{createTrustManager} receive null inputs and raise \texttt{NullPointerException}s wrapped in higher-level exceptions.

\textit{\textbf{The final LLM-enhanced report correctly identifies the root cause}}, which is the absence of required properties such as \texttt{keyStoreLocation} or \texttt{keyStorePassword}. It pinpoints the relevant methods and classes involved in the failure and suggests an actionable fix, initializing the system properties before instantiating the authentication provider. This aligns precisely with the developer patch, which added null checks in the \texttt{X509AuthenticationProvider} constructor to guard against misconfiguration. 
This example illustrates how LLM-based enhancement can structure and enrich crash reports by (i) linking stack traces to implicated methods, (ii) clarifying likely root causes, and (iii) suggesting actionable fixes. The two instantiations differ only in how they collect evidence: \textsc{Agentic-LLM} obtains an expanded evidence bundle via iterative, graph-guided traversal, whereas \textsc{Direct-LLM} supplies only the stack trace referenced methods; we empirically compare their effectiveness next using quantitative metrics, manual judgment, and a user study.

\section{Experiment Results}
\label{sec:discussion}

In this section, we present the results of our research questions (RQs). Each RQ is composed of the motivation, approach, and results. 


\phead{\textbf{Implementation and Environment.}}  
\label{sec:implementation_environment}
We use \texttt{javalang}~\cite{javalang}, a Java source code parser, to extract method-level representations from Java source code. These representations are further used to construct method-level call graphs, which provide the agent with structured context for reasoning about inter-method relationships. 
We use OpenAI's \texttt{gpt-4o-mini-2024-07-18} model, a more cost-effective yet capable LLM~\cite{gpt-4o-mini}, as the underlying LLM to reduce costs. 
To improve consistency and reduce response variation, we set the temperature parameter to 0 during model inference~\cite{holtzman2019curious}. We use LangChain~\cite{Chase_LangChain_2022}, a framework managing LLM agents and tools, for our agent implementation.


\subsection*{RQ1: Do LLM-enhanced crash reports improve fault localization and possible fix accuracy?}
\label{RQ1}
\phead{Motivation.}
As discussed in Section \ref{sec:preliminary_analysis}, original developer-written crash reports (DevCR) frequently omit critical diagnostic information, hindering their usefulness for fault localization and fixing. Among the various diagnostic fields, problem location and possible fix are the only two fields with ground truth (i.e., the set of modified methods in the fixing commit and the developer patch) that allow large-scale and automated evaluation. 
Hence, in this RQ, we compare \devcrllm\ and \tooltable\ on their effectiveness for fault localization (accuracy of predicted problem location) and possible fix (similarity to developer patch), and whether enhanced reports provide more accurate diagnostic information than DevCR.

\phead{Approach.}
\underline{\textit{Problem Location}}. For \devcrllm\ and \tooltable, we directly use the report’s \texttt{problem\_location} field, which lists candidate \textbf{\textit{faulty methods}}. We compare this candidate list with the ground truth list of methods that developers actually modified to fix the crash. Similar to prior studies~\cite{sohn2017fluccs, lou2021boosting, li2019deepfl, kang2024quantitative, qin2024agentfl}, if one match is found, the report is counted as correctly localized. 
We report the problem localization accuracy as the proportion of crash reports, out of all reports in that system, where the problem is correctly localized.
For DevCR, since location information is typically absent, we follow prior informational retrieval-based fault localization work~\cite{razzaq2021boostnsift}. We apply BM25~\cite{robertson1995okapi}, which has shown great localization accuracy~\cite{razzaq2021boostnsift}, using the entire crash report as the query to find the most relevant methods in the code. 
We evaluate DevCR using Recall at Top-N, which measures the percentage of crash reports for which at least one faulty method appears within the top-N ranked results. Following prior work~\cite{parnin2011automated}, we report Top-1, Top-3, and Top-5. 


\underline{\textit{Possible Fix}}. We evaluate the \texttt{possible\_fix} field from \devcrllm \ and \tooltable. We assess the fix accuracy by comparing the possible fix code with the ground truth fixed code in the developer patch, computing CodeBLEU~\cite{ren2020codebleu} between the two. CodeBLEU captures both syntactic and semantic similarity, so possible fixes with higher scores are more likely to align with developer-intended changes. For each system, we report the mean CodeBLEU score computed over crash reports that successfully localized the problem. 
We excluded DevCR in the analysis since, as shown in Section~\ref{sec:preliminary_analysis}, only about 20\% of developer-written crash reports include a possible fix.

\begin{table}[t]\centering
\caption{Problem localization accuracy (percentage) of Developer-written crash report (DevCR) and LLM-enhanced crash reports (\devcrllm\ and \tooltable).}\label{tab: problem_location}
\scalebox{.7}{
\begin{tabular}{lcccccc}\toprule
&\multirow{2}{*}{\textbf{\devcrllm}} &\multirow{2}{*}{\textbf{\tooltable}} &\multicolumn{3}{c}{\textbf{DevCR}} \\\cmidrule{4-6}
& & &\textbf{ top-1} &\textbf{ top-3} &\textbf{ top-5} \\\midrule
ActiveMQ &25.71 &\textbf{34.29} &5.71 &11.43 &20 \\
Hadoop Common &44.74 &\textbf{55.26} &10.53 &15.79 &23.68 \\
HDFS &42.47 &\textbf{45.21} &8.22 &16.44 &23.29 \\
MapReduce &35.38 &\textbf{36.92} &6.15 &9.23 &13.85 \\
YARN &47.01 &\textbf{48.72} &12.82 &23.08 &29.06 \\
Hive &\textbf{41.67} &\textbf{41.67} &14.17 &20 &25 \\
Storm &30.56 &\textbf{36.11} &8.33 &11.11 &13.89 \\
ZooKeeper &\textbf{25} &\textbf{25} &12.5 &\textbf{25} &\textbf{25} \\ \hline
Average &40.24 &\textbf{43.09} &10.57 &17.28 &22.97 \\
\bottomrule
\end{tabular}}
   \begin{tablenotes}
     \item \footnotesize{\textbf{Note:} The best accuracy of each system is highlighted in \textbf{bold}.} 
   \end{tablenotes}
\end{table}

\phead{Results.}
\noindent\textbf{\em LLM-enhanced crash reports substantially outperform DevCR by 400\% for problem localization.} Table~\ref{tab: problem_location} reports the problem localization accuracy. Across eight studied systems, DevCR achieves 10.57\% (Top-1), 17.28\% (Top-3), and 22.97\% (Top-5), whereas \tooltable\ and \devcrllm\ localize 43.09\% (212/492) and 40.24\% (198/492), both showing around 4 times improvement over DevCR's Top 1. 
These gains arise from (i) structured fields (\texttt{problem\_location}) that turn noisy information into explicit candidate methods, reducing ambiguity when matching to ground truth, and (ii) richer context, as enhanced reports incorporate code-aware evidence (e.g., method bodies aligned to stack traces), enabling more semantically accurate localization than DevCR, whose information often lacks key details.

\noindent\textbf{\em \tooltable\ outperforms \devcrllm\ on problem localization task, largely due to richer code context.} 
Averaged across systems, \tooltable\ achieves 43.09\% localization accuracy compared with 40.24\% for \devcrllm, achieving a 7.01\% improvement, and leads in 6 of 8 systems (notably Hadoop Common 55.26\% compared with 44.74\%, and Storm 36.11\% compared with 30.56\%). 

In total correct localizations, \tooltable\ and \devcrllm\ overlap on 169 crashes, about 85\% of \devcrllm’s and 80\% of \tooltable’s results. The large overlap suggests both enhancement modes capture many of the same issues. We further examine the differences. 
\devcrllm\ succeeds in some cases where the crash is directly related to the immediate stack trace frames, but its single-shot design may miss deeper causes. In some cases, the agentic design may hallucinate and miss the faulty method. Conversely, \tooltable’s iterative traversal uncovers faults that require exploring beyond the stack frames, finding the faulty method for more complicated crashes. 

\begin{wraptable}{r}{0.55\textwidth} 
\caption{
Average CodeBLEU (\%) between LLM generated \texttt{possible fix} and the developer patch for matched method cases, by project.
}
\centering
\scalebox{.7}{
\setlength{\tabcolsep}{0.61cm}
\begin{tabular}{lcc}
    \toprule
    \textbf{Project} & \textbf{\devcrllm} & \textbf{\tooltable} \\ 
    \midrule
    ActiveMQ          & 58.63 & 66.40 \\
    Hadoop Common     & 49.91 & 50.79 \\
    HDFS              & 59.41 & 58.68 \\
    MapReduce         & 48.11 & 51.28 \\
    YARN              & 53.28 & 56.81 \\
    Hive              & 64.72 & 65.30 \\
    Storm             & 54.69 & 42.44 \\
    ZooKeeper         & 0.00 & 28.24 \\
    \midrule
    \textbf{Overall Average} & \textbf{56.51} & \textbf{57.28} \\ 
    \bottomrule
\end{tabular}
}\vspace{-0.2cm}
\label{tab:codebleu_comparison}
\end{wraptable}

Table~\ref{tab:codebleu_comparison} shows mean CodeBLEU scores for matched cases. Overall, the \textit{\textbf{possible fixes from enhanced reports align well with developer patches}}. The overall average CodeBLEU is 57.28\% for \tooltable\ and 56.51\% for \devcrllm, with most projects above 50\% (e.g., ActiveMQ 66.40\%, Hive 65.30\%, YARN 56.81\%). \tooltable\ yields slightly higher CodeBLEU on average, while \devcrllm\ remains competitive in some projects.

We also examine the monetary costs of the two LLM-enhanced crash reports. The \textit{\textbf{per-report cost is approximately \$0.01 for \tooltable\ and \$0.005 for \devcrllm}}, both modest compared to the potential developer time saved. This tradeoff suggests that \tooltable\ is most suitable for high-severity crashes and where the cause may be more complicated, while \devcrllm\ offers a cost-effective option for large-scale or budget-sensitive settings. Nevertheless, both still provide substantial improvements over DevCR.




\rqboxc{
LLM-enhanced crash reports substantially outperform DevCR, raising Top-1 localization accuracy from 10.6\% to 40.2-43.1\%. The possible fixes also have a high CodeBLEU score (56-57\%) with the ground truth patch. \tooltable performs slightly better than \devcrllm on both tasks.}

\subsection*{RQ2: Do LLM-enhanced crash reports improve diagnostic quality?}
\label{RQ2}


\phead{Motivation.} 
While RQ1 quantifies correctness using ground truth measures (problem location and possible fix), it does not capture the qualitative aspects of report content. Many critical dimensions of diagnostic quality, such as the root cause explanation, require human judgment and are difficult to evaluate automatically. 
In this RQ, we perform a manual evaluation on the three types of reports (DevCR, \devcrllm, and \tooltable) along three dimensions: location clarity, root-cause explanation, and repair guidance.



\phead{Approach.}  
\underline{\textit{Manual Study}}.
For each crash, we manually compare three report variants across three diagnostic dimensions: \textbf{Crash Location}, \textbf{Crash Root Cause}, and \textbf{Crash Repair}.
Similar to prior studies \cite{kang2024quantitative, rafi2025revisiting}, the first two authors of this paper (i.e., A1 and A2) conducted a manual study following four phases:
(1) We applied stratified random sampling \cite{baltes2022sampling} to select 100 crashes for manual study. We chose this number due to the extensive effort needed to investigate the crashes. 
For each crash, we have three variants of the crash report, totaling 300 reports to be inspected along with the corresponding code and fixes. 
(2) A1 drafted the definition of quality levels for each dimension. A1 and A2 collaboratively labelled an initial set of 50 randomly sampled reports using these draft levels and refined the level definitions during this phase.
(3) A1 and A2 independently labelled each dimension of the rest of the 250 sampled reports with quality levels defined in Phase 2. There is no new category derived in this phase.
(4) A1 and A2 then compared their results across three dimensions for each crash report until reaching a consensus. The results in this phase have a Cohen’s kappa of 0.9568, indicating an almost perfect agreement \cite{cohen1960coefficient}.

\underline{\textit{LLM-as-a-Judge}}.
We complement the manual study with an \emph{LLM-as-a-Judge} evaluation to cover all 492 crashes. 
LLM-as-a-Judge means that LLMs emulate human reasoning and evaluate specific inputs against a set of predefined rules. Recent studies \cite{gu2024survey, zhang2024large, ahmed2024can} have shown that LLM reaches a similar agreement level as that of humans. 
Concretely, we provided the quality-level definitions derived from the manual study as explicit criteria in the prompt for each dimension, the report, associated source code, and ground truth. We then asked the LLM to assign a quality level to every report and to compare the three variants under the same crash.

\begin{table*}
\scriptsize
\centering
\caption{Evaluation results: Manual evaluation across 100 crashes and LLM-as-a-Judge evaluation across 492 crashes.}

\label{tab:merged-eval}

\setlength{\tabcolsep}{3.2pt} 
\renewcommand{\arraystretch}{0.9} 

\scalebox{0.87}{
\begin{tabular}{p{1.3cm}p{4.3cm}lll lll}
\toprule
\textbf{Dimension} & \textbf{Diagnostic Quality Levels} 
& \multicolumn{3}{c}{\textbf{Human Evaluation (100)}} 
& \multicolumn{3}{c}{\textbf{LLM-as-a-Judge (492)}} \\
\cmidrule(lr){3-5} \cmidrule(lr){6-8}
& & \textbf{DevCR} & \textbf{\devcrllm} & \textbf{\tooltable}
  & \textbf{DevCR} & \textbf{\devcrllm} & \textbf{\tooltable} \\
\midrule

\multirow{3}{*}{\makecell[l]{\textbf{Crash}\\ \textbf{Location}}} 
& Crash-causing \textbf{method} mentioned 
& 42 (42\%) & 100 (100\%) & 100 (100\%) 
& 165 (33.54\%) & 490 (99.59\%) & 492 (100\%) \\
& Crash-causing \textbf{class} mentioned 
& 13 (13\%) & 0 (0\%) & 0 (0\%) 
& 10 (2.03\%) & 0 (0\%) & 0 (0\%) \\
& No mention of crash location 
& 45 (45\%) & 0 (0\%) & 0 (0\%) 
& 317 (64.43\%) & 2 (0.41\%) & 0 (0\%) \\

\midrule

\multirow{5}{*}{\makecell[l]{\textbf{Crash}\\ \textbf{Root Cause}}} 
& \textbf{Complete root cause} of crash provided  
& 30 (30\%) & 46 (46\%) & 58 (58\%) 
& 113 (22.97\%) & 169 (34.35\%) & 173 (35.16\%) \\
& Incomplete root cause 
& 70 (70\%) & 54 (54\%) & 42 (42\%) 
& 379 (77.03\%) & 323 (65.65\%) & 319 (64.84\%) \\
& \SFviii Proximate cause w/ stack traces 
& \SFviii 7 (7\%) & \SFviii 54 (54\%) & \SFviii 42 (42\%) 
& \SFviii 66 (13.41\%) & \SFviii 323 (65.65\%) & \SFviii 319 (64.84\%) \\
& \SFviii Trigger/environment w/ stack traces 
& \SFviii 51 (51\%) & \SFviii 0 (0\%) & \SFviii 0 (0\%) 
& \SFviii 292 (59.35\%) & \SFviii 0 (0\%) & \SFviii 0 (0\%) \\
& \SFii Stack traces only 
& \SFii 12 (12\%) & \SFii 0 (0\%) & \SFii 0 (0\%) 
& \SFii 21 (4.27\%) & \SFii 0 (0\%) & \SFii 0 (0\%) \\

\midrule

\multirow{3}{*}{\makecell[l]{\textbf{Crash}\\ \textbf{Repair}}} 
& \textbf{Ground truth} repairs provided 
& 19 (19\%) & 29 (29\%) & 41 (41\%) 
& 47 (9.55\%) & 134 (27.24\%) & 149 (30.28\%) \\
& \textbf{Alternative} repairs provided 
& 3 (3\%) & 71 (71\%) & 59 (59\%) 
& 39 (7.93\%) & 353 (71.75\%) & 343 (69.72\%) \\
& No repair suggestion 
& 78 (78\%) & 0 (0\%) & 0 (0\%) 
& 406 (82.52\%) & 5 (1.02\%) & 0 (0\%) \\
\bottomrule
\end{tabular}
}
\end{table*}

\phead{Results.}
Table~\ref{tab:merged-eval} shows the diagnostic quality levels that we manually derived and the study results across 100 crash reports. 
For crash location, \textit{\textbf{both \devcrllm\ and \tooltable\ correctly identify the crash-causing method in every case (100\%)}}, while DevCR mentions the method in only 42\% of reports, the class in 13\%, and omits the location entirely in 45\%. Note that the crash location is the location where the crash occurred, and not the actual fix location. The information may be embedded in the stack traces and the associated source code, yet it is often implicit or ambiguous in DevCR, forcing developers to infer the relevant method themselves. On the other hand, the enhanced reports explicitly extract and structure this information, making the crash location immediately clear and actionable. 

For the root cause, we analyzed the report to see if it describes the primary cause of the crash (i.e., what caused the failure). 
For example, consider a \texttt{NullPointerException}: a \textit{complete} explanation would name the symptom and explain why the object became null (for example, \textit{A NullPointerException occurred because the object was never initialized under certain startup conditions; the fault lies in the initialization path, so the initialization should be added or guarded}). By contrast, a \textit{proximate} or partial explanation merely restates the symptom (for example, \textit{A NullPointerException occurred because the program attempted to use a null object.}) without identifying any reason.

DevCR provides a complete explanation of the cause of the crash in only 30\% of cases, with most reports limited to partial triggers (51\%), where only the conditions or issue behavior are provided, but the actual reason is not explicitly mentioned. For example, a partial-trigger report might say \textit{the server crashes on startup when SSL is enabled} or \textit{a NullPointerException occurs only when uploading large files}, which tells when the crash happens but not why it happens. Another 12\% contain only raw stack traces.
In contrast, \devcrllm\ provides the complete root cause to 46\% and \tooltable\ further to 58\%. 
While both enhanced crash reports provide some reasons for the cause (i.e., no trigger/environment-only and stack trace only), they differ in depth: \textit{\textbf{\devcrllm\ often stops at proximate causes (54\%), whereas \tooltable\ provides the complete root cause for more crashes}}. 
This highlights the benefit of agentic iterative reasoning for producing better explanations.

For crash repair, DevCR rarely provided actionable fixes. Only 22\% of the crash reports contained some repair suggestion, with 19\% offering ground truth fixes (i.e., identical to developer fixes). In contrast, both LLM-enhanced approaches provide either alternative or {ground truth} repairs. 
An alternative repair does not match the developer's patch, but would plausibly resolve the failure. For example, for a \textit{NullPointerException caused by missing initialization}, a ground truth suggestion would add the same initialization. In contrast, an alternative might add a null check at the call site to avoid the crash. This prevents the crash but follows a different remediation strategy. 
\textit{\textbf{\devcrllm\ produced alternative repairs in 71\% of reports and {ground truth} fixes in 29\%, while \tooltable\ achieved stronger results, with 59\% alternative repairs and 41\% {ground truth} fixes}}. These findings indicate that LLM-enhanced reports not only supply more consistent repair guidance but also generate higher-quality fixes than developer-written reports.

To conduct a large-scale automated analysis using LLM, we first apply LLM-as-a-Judge on the 100 manually studied crash reports and cross-check with our manual labels. The result shows that LLM's results align closely with manual labels (with a Cohen’s kappa of 0.7728, indicating substantial agreement~\cite{cohen1960coefficient}). Similar to prior studies~\cite{gu2024survey, zhang2024large, ahmed2024can}, the result supports the reliability of the large-scale evaluation by LLM. 
Table~\ref{tab:merged-eval} (columns 6-8) presents the diagnostic quality evaluation results of LLM-as-a-Judge on all reports of 492 crashes. We observe the same trend as in the manual study. DevCR often omitted method locations (33.5\%), gave incomplete root causes (77\%), and rarely suggested repairs (82.5\%). In contrast, \textit{\textbf{both LLM-enhanced approaches consistently identified the crash-causing method, provided better root cause reasoning, and suggested better fixes}}. \tooltable\ showed a slight improvement over Direct-LLM, providing more complete root cause analysis (35.16\% vs 34.35\%) and ground truth repairs (30.28\% vs 27.24\%). These results highlight that LLM-enhanced reports improve all three dimensions of diagnostic quality, and \tooltable\ provides better results compared to \devcrllm.




\rqboxc{Both manual and LLM-as-a-Judge evaluations show that LLM-enhanced reports outperform DevCR in crash location, root cause, and repair. \tooltable\ provides slightly better results than \devcrllm\ on all dimensions.}

\subsection*{RQ3: Do developers perceive LLM-enhanced crash reports as more useful than developer-written crash reports?}
\label{RQ3}

\phead{Motivation.}  
While LLM-enhanced crash reports may appear more structured and comprehensive, it is essential to determine whether these enhancements are genuinely useful in real-world debugging. In this RQ, we conduct questionnaires with developers to determine whether such enhanced crash reports are more helpful than the developer-written ones. 

\phead{Approach.}  
We conducted a user study to assess the practical usefulness of LLM-enhanced crash reports. Our study involved 16 participants, comprising seven graduate students and nine software developers with strong programming backgrounds. 11 participants reported having more than 5 years of programming experience, and five reported having 3–5 years. All 16 participants listed Python as a primary language, while 10 also reported Java (with multiple selections allowed). 


We designed the user study to reflect real-world debugging. The participants were asked to analyze crash reports and corresponding source code to understand and debug the issues. Due to the extended time and significant effort needed for the user study, each participant received four distinct crash reports: two developer-written reports (DevCR) and two LLM-enhanced reports generated by \textsc{Agentic-LLM}. We use the reports generated by \tooltable\ instead of \devcrllm\ because our evaluations in RQ1 and RQ2 show that the agentic approach consistently produces higher-quality reports.
On average, participants spent 20 minutes
per crash (about 80 minutes in total for four reports), reflecting the effort required to locate the
problem in the code, identify the root cause, and propose a fix. 
Participants were not told which reports were enhanced, and we only revealed whether LLM enhanced the report after they completed a debugging task. 

To ensure diversity, we selected six real-world crash scenarios covering five exception types commonly found in our dataset: \emph{Null Pointer Exception (NPE)}, \emph{Array Index Out of Bound Exception}, \emph{IO Exception}, \emph{Channel Exception}, and \emph{Illegal Argument Exception}. As nearly 20\% of our corpus involved Null Pointer Exceptions, we included two such crashes (one developer-written, one LLM-enhanced). We distributed them across groups such that each participant received only one NPE report (either developer-written or enhanced), along with three additional non-NPE crashes. 
In total, this yielded 32 evaluations for developer-written reports and 32 for LLM-enhanced reports.


After each debugging task, participants answered four 5-point Likert-scale~\cite{joshi2015likert} questions on the report’s helpfulness for (i) locating the problem, (ii) understanding the root cause, (iii) suggesting a fix, and (iv) overall quality, with the option of free-text explanations. An "\textit{I don’t know}" option was also available but rarely used: five times for possible fix, once for root cause, and once for problem location, indicating participants were generally able to form judgments on the assigned tasks.

\phead{Results.}  
\noindent\textbf{\textit{Participants overwhelmingly preferred LLM-enhanced reports, rating them as significantly more helpful than original reports across all debugging dimensions.}} 
Table~\ref{tab:user-study-results} presents the average participant ratings on LLM-enhanced reports and DevCRs.
Participants consistently preferred LLM-enhanced crash reports to the DevCRs across all debugging dimensions. Helpfulness for locating the problem rose from 3.38 to 4.69, root-cause explanations from 3.16 to 4.66, fix suggestions from 2.66 to 4.53, and overall report quality from 3.00 to 4.59. These improvements indicate that reports enhanced by LLMs with repository context, coupled with iterative agentic reasoning, providing a clearer guidance for where the fault, why the failure occurs, and how to repair it.
The standard deviation of ratings of LLM-enhanced reports in each evaluation dimension was smaller than that of DevCRs, which reflects the quality variants between reports were also narrowed after LLM enhancement. 
Overall, the user study shows that developers perceive LLM-enhanced reports provide more stability and higher quality than DevCRs, with the most noticeable improvement in the actionability of fix suggestions.

\begin{wraptable}{r}{0.55\textwidth} 
\caption{Mean participant ratings on a 1-5 scale (5 = best, 1 = worst) comparing developer-written vs. LLM-enhanced crash reports, where the numbers in the parentheses represent the standard deviation (SD).}
\label{tab:user-study-results}
\centering
\scalebox{0.72}{
\setlength{\tabcolsep}{0.35em}
\begin{tabular}{lccc}
\toprule
\textbf{Evaluation Dimensions} & \textbf{DevCR} & \textbf{\textsc{LLM-enhanced}} \\
\midrule
Helpfulness in finding problem location      & 3.38 (0.87) & 4.69 (0.47) \\
Helpfulness in explaining root cause         & 3.16 (1.11) & 4.66 (0.55) \\
Helpfulness in suggesting a fix              & 2.66 (1.21) & 4.53 (0.72) \\
Overall report quality                       & 3.00 (0.95) & 4.59 (0.56) \\
\bottomrule
\end{tabular}
}
\vspace{-0.3cm}
\end{wraptable}

Most participants described LLM-enhanced reports as \textit{well-structured}, \textit{clear}, and \textit{easy to follow}. Several noted they were \textit{more than enough} or \textit{very helpful in solving the issue}. One participant mentioned: \textit{This crash report is high quality. It not only provided almost all details about this crash, but also gave me a useful suggestion for fixing it.} Another said: \textit{The format is structured and easy to find the crash, and it also gives a fix suggestion.}
Participants appreciated the structured format and inclusion of context and fix rationale. There are also suggestions for improvement, such as including test cases or input examples. 

In contrast, most participants found the original developer-written crash reports \textit{vague} or \textit{lacking diagnostic context}. 
One participant noted: \textit{It should show the problem rather than just the error message, which is hard to debug.} Others commented on missing details such as input conditions or environment information: \textit{More details and test cases could have eased the crash trace}, and \textit{The stack trace shows the location but not why it failed.}
One participant even suggested: \textit{A fixed template for developers could help match the structure of LLM reports.}

Overall, participant feedback highlights that LLM-enhanced reports not only improve the clarity of crash reports but also enhance their diagnostic value, ultimately making debugging more efficient and accessible.

\rqboxc{Developers found LLM-enhanced crash reports significantly more useful than developer-written crash reports across all evaluation aspects, particularly for identifying fault locations, understanding root causes, and suggesting actionable fixes. These findings support the potential of LLM-based enhancement for real-world debugging.}

\section{Threats to Validity}
\label{sec:threats}

\noindent\textbf{Internal Validity.} 
Our methodology relies on several components whose correctness may influence the outcome of the study. Specifically, the effectiveness of our LLM-enhanced crash reports depends on the quality of inputs such as stack traces, source code at the reported revision, and the developer-written reports. If these inputs are incomplete, noisy, or incorrectly extracted (e.g., due to parsing errors or repository checkout issues), the generated report may fail to reflect the true context of the crash. We mitigate this by performing validation checks during data extraction and using verified source snapshots from version-controlled repositories.

\noindent\textbf{External Validity.}  
To conduct our study, we use 492 crash reports from eight widely used open-source Java systems. While these projects span diverse domains and codebases, our findings may not generalize to software written in other programming languages (e.g., C++, Python) or to industrial systems with different development and reporting practices. Additionally, our analysis focuses on crash reports containing stack traces; generalization to non-crash-related bug reports is not evaluated. Future work is needed to extend the methodology to broader software ecosystems and bug types.

\noindent\textbf{Construct Validity.}  
We measure the effectiveness of enhanced crash reports quantitatively by evaluating problem-location accuracy, possible fix similarity (CodeBLEU), and IR-based fault localization. To complement these automated metrics, we conduct a manual study on a stratified sample of 100 crashes and scale the evaluation across all reports using an LLM-based judge. While these measures capture both quantitative and qualitative aspects of report usefulness, they may not cover every dimension of practical debugging.


\section{Conclusion}\label{sec:conclusion}

Crash reports are a critical part of the software maintenance workflow, but our analysis of 492 real-world reports shows that many are incomplete and lack the diagnostic detail needed to support efficient debugging. In this paper, we investigated whether LLM-based enhancement can improve crash reports by enriching them with fault locations, root-cause explanations, and fix suggestions. 
We compared two enhancement strategies: a single-shot approach (\devcrllm) and an iterative, agentic workflow (\tooltable). Our results show that LLM-enhanced crash reports substantially improve problem localization accuracy (raising Top-1 accuracy from 10.6\% for DevCR to 40.2–43.1\%) and provide high-quality possible fixes (CodeBLEU is around 56–57\% compared to developer patch). Our manual and LLM-as-a-Judge evaluations further show that enhanced reports provide clearer root-cause explanations and better repair guidance. In particular, \tooltable\ provides better diagnosis information compared to \devcrllm\ because of its deeper analysis on the code across the repository. Finally, our user study with 16 participants shows that enhanced reports improve developers’ ability to understand and resolve crashes, with the greatest improvement in repair guidance, compared to original developer-written reports. 
Overall, our findings show that LLM-enhanced crash reports provide richer and more actionable diagnosis information during debugging. 

\balance
\bibliographystyle{ACM-Reference-Format}
\bibliography{reference}

\end{document}